\documentstyle[12pt,epsf]{article}

\renewcommand{\a}{\alpha}
\renewcommand{\b}{\beta}

\renewcommand{\d}{\delta}

\newcommand{\cD}{{\cal D}}

\newcommand{\ep}{\epsilon}

\newcommand{\ve}{\varepsilon}
\newcommand{\si}{\sigma}   
\newcommand{\Si}{\Sigma}
\newcommand{\th}{\theta}

\newcommand{\p}{\partial}

\newcommand{\vs}{\vskip.5cm}

\newcommand{\lbr}{\left[}
\newcommand{\rbr}{\right]}
\newcommand{\lan}{\langle}
\newcommand{\ran}{\rangle}

\newcommand{\beq}{\begin{equation}}
\newcommand{\eeq}{\end{equation}}
\newcommand{\beqa}{\begin{eqnarray}}
\newcommand{\eeqa}{\end{eqnarray}}

\newcommand{\prd}[1]{{ \it Phys.~Rev.}~{\bf D{#1}}}
\newcommand{\prl}[1]{{ \it Phys.~Rev.~Lett.}~{\bf {#1}}}
\newcommand{\plb}[1]{{ \it Phys.~Lett.}~{\bf {#1B}}}
\newcommand{\pla}[1]{{ \it Phys.~Lett.}~{\bf {#1A}}}
\newcommand{\npb}[1]{{ \it Nucl.~Phys.}~{\bf B{#1}}}

\newcommand{\rmp}[1]{{ \it Rev.~Mod.~Phys.}~{\bf {#1}}}
\newcommand{\cmp}[1]{{\it Comm.~Math.~Phys}~{\bf {#1}}}

\newcommand{\mpl}[1]{{\it Mod.~Phys.~Lett}~{\bf {#1}A}}
\newcommand{\anm}[1]{{\it  Ann.~Math.}~{\bf {#1}}}

\newcommand{\Xd}{\dot X}
\newcommand{\Yd}{\dot Y}
\newcommand{\Xdd}{\ddot X}
\newcommand{\Ydd}{\ddot Y}

\newcommand{\vn}{{\vec n}}
\def\half{{\mbox{\small  $\frac{1}{2}$}}}
\newcommand{\postscript}[2]
 {\setlength{\epsfxsize}{#2\hsize}
  \centerline{\epsfbox{#1}}}
\def\llap#1{\hbox to 0pt{\hss#1}}
\def\pola{a\llap{\hbox{\char'30\kern-1.2pt}}}
\def\pole{e\llap{\hbox{\char'30\kern-.8pt}}}

\begin{document}
\baselineskip 22pt plus 2pt

\begin{titlepage}
\renewcommand{\thefootnote}{\fnsymbol{footnote}}
\begin{flushright}
\parbox{1.in}
{
 IFT-24/94\\
}
\end{flushright}
\vspace*{1.5in}
\begin{centering}
{\Large IMMERSIONS AND FOLDS\\ \vskip.3cm IN \\ \vskip.3cm STRING THEORIES OF
GAUGE FIELDS \footnote{ 
Work supported, in part, 
by Polish State Committee for Scientific Research (KBN).
}}\\
\vspace{2cm}
{\large        Jacek Pawe\l czyk}\\
\vspace{.5cm}
        {\sl Institute of Theoretical Physics, Warsaw University,\\
        Ho\.{z}a 69, PL-00-681 Warsaw, Poland.}\\
\vspace{.5in}
\end{centering}
\begin{abstract}
 A two-dimensional string model with dynamical cancellation of folds is 
considered. The action of the model contains the self-intersection number
which is defined for  surfaces immersed into  4D
targets. The two additional variables are not dynamical and live on a compact
manifold. In this sense the model is a compactification of 
a 4D theory. The cancellation forces the string $\th$ angle to be equal 
$\pi$. Candidates for string states are constructed. Some mathematical 
background is given.

\end{abstract}

\end{titlepage}

\setcounter{section}{-1}
\setcounter{footnote}{0}
\renewcommand{\thefootnote}{\alph{footnote}}

\section{Introduction}
\label{sec:intro}

It is strongly believed that dynamics of gauge fields can be described
in terms of a string theory. The idea was supported by the lattice strong  
coupling expansion \cite{lat} and the $1/N_c$  expansion \cite{thooft}. The
latter applied  
in 2-dimensional (2D) models gave several well established relations
between QCD$_2$ (or YM$_2$) and a string theory \cite{bars,gross}. It
appeared that the crucial role 
is played by  the no fold condition, which says that surface-to-surface maps
with folds do 
not contribute to gauge theory functional integrals. This 
restricts  the set  of relevant maps, defining a 2D string theory, to a
residual set. Moreover, the 
results indicate that the proper string action
should contain 
the Nambu-Goto term.  It is well known that the Nambu-Goto
term alone can not give the correct 
picture, because the appropriate functional integral can not suppress
folds.

Recently two solutions to the
problem of folds  have  been proposed \cite{moore,horava}. 
The idea is to construct a 
topological field theory localized on an appropriate space of maps without
folds.
In the first work it is the space of holomorphic maps, which corresponds to 
one chiral sector \cite{gross}. The another chiral sector is given by 
anti-holomorphic maps. For completeness of the approach both spaces should be
compactified  and glued together.
This  proposal has been almost completely worked out yielding
results in accordance with YM$_2$ \cite{gross}, at least wherever the calculations 
has been  finished. 
Another proposition deals with the space of harmonic
maps \cite{horava}. In a sense,
it is  more natural because holomorphic and anti-holomorphic maps are subsets
of harmonic maps. Unfortunately it has not been worked out
to this extend as the previous proposal.
Despite these successes many properties of string picture of gauge fields are
still missing e.g. it is not clear what is the correct 4D model.

In Ref.7 a string theory model with a dynamical cancellation of
folds 
has been proposed.
The model consists of two terms: the Nambu-Goto
action and  a topological term, the self-intersection number \cite{whitneysi}. 
The latter is well defined in a target space of dimension 4, thus
two additional variables with values in a 2D space (farther called
vertical) have been introduced. 
The model
properly suppresses folded configurations yielding null partition function
and null transition amplitudes for 
microscopic states (infinitesimal punctures) for string propagating on the flat
2D target space-time.

This paper is a continuation of Ref.7. It purpose it two-fold: 
we want to generalize results in different directions and
elaborate on a 1D example, which illustrates most of the basic ideas. We also
include some mathematical background. The paper is  
organized as follows:
in Sec.\ref{sec:r2r2} we describe surface-to-surface maps with special emphasis on some
peculiarities of folds and cusps. Section \ref{sec:imm} summary facts concerning
immersions of surfaces in 4D spaces. 
In the next section we dwell upon a 1D example of non-backtracking particle. 
We discuss the functional measure and the propagator on a circle $S^1$.
Section \ref{sec:string} contains the
main results of the paper. In first part we generalize results of  
Ref.7 for an arbitrary vertical (orientable) manifold. 
Then a compactification on $S^2$ is considered. We show that it forces
$\theta=\pi$.  Next we discuss shortly YM$_2$ on compact Riemannian surfaces.
In final subsection we go to 4D case and
construct string states which bear resemblance with YM$_4$ states.
We also give there a
general formula for the self-intersection number of a surface with boundary. 
The last section is devoted to  cancellation of singular
contributions for 3D string theories. 

\section{Surface-to-surface maps.}
\label{sec:r2r2}

Generic surface to surface maps and their singularities were classified by
Whitney \cite{r2r2}. Below we briefly summary his results.

Let $X:\Si \to M$ denote a smooth map of a  surfaces $\Si$ in a 2D manifold
$M$ and $S_1(X)=\{p\in \Si: {\rm rank}(dX(p))=1\}$ its set of singular points. 
In a local coordinate patch the condition means that
the matrix of the induced metric $g_{ab}=\p_a X^\mu \p_b X^\mu$ 
has one vanishing eigenvalue.
For generic maps $S_1$ is a 1D submanifold of $\Si$.
In particular, if $\Si$ is a closed surface (compact, connected, 
without boundary) then $S_1$ is a finite family of disjoint curves (loops)
embedded in $\Si$.

Farther we define $S_{1,1}=\{p\in S_1:
{\rm rank}(d(X|_{S_1}(p))=0\}$ i.e. $S_{1,1}$ is a set of (disjoint) points
at which eigenvector corresponding to zero eigenvalue is tangent to  $S_1$.
In a local coordinate system (simple) folds and
cusps have the 
following form: folds, $(X^1,X^2)=(s^2,t)$; cusps,  $(X^1,X^2)=(st-s^3,t)$.

Let $N$ be a neighborhood of $S_1$. $N$ may have topology of a cylinder or a
M{\"o}bius strip. 
The second case can happen for  non-orientable
manifolds. One can think of N as a (disc) bundle over $S_1$.
Let  $K$ be a line bundle over $S_1$ with fibers given by the
kernel of $dX$. This bundle can also be orientable (a cylinder) or
non-orientable (a M{\"o}bius strip). By $w(N),\; w(K)$ we denote the first
Steifel-Whitney classes of the appropriate bundles. 
Both Stiefel-Whitney classes are in $H^1(S_1,Z_2)$.
Then, of course $w=0$ for the orientable bundle
and $w\neq 0$ ( equals to the only non-trivial element of 
$H^1(S_1,Z_2)$) for the  non-orientable bundle. 
Finally we define the first
Steifel-Whitney class of $S_{1,1}$, $w(S_{1,1})$, to be the class 
in $H^1(S_1,Z_2)$ Poincare dual to the homology class represented by
$S_{1,1}$. By this definition $w(S_{1,1})$ is zero if the number of points
in $S_{1,1}$ is even and it equals to the only non-trivial element of 
$H^1(S_1,Z_2)$ otherwise.
Here we recall that addition of two non-trivial elements of $H^1(S_1,Z_2)$
gives the trivial element  due to the mod 2 property.

There is an nice relation between $w(N)$, $w(K)$ and $w(S_{1,1})$
 \cite{blank}: $w(N)+w(K)+ w(S_{1,1})=0$. It says that for orientable $\Si$
(i.e. $w(N)=0$) and orientable $K$ there are even number of cusps, while for
non-orientable $K$ odd number of cusps.  
In the next section 
we
shall be mainly interested in the $K$ bundle thus the information whether it is
orientable or not will be of some importance. 

We know that generic maps do not contribute to
the partition function \cite{gross}. 
It means that any string representation of  YM$_2$ have to have  built in
cancellation of these maps.

\section{Immersions}
\label{sec:imm}

In the previous section we have described the space of generic
surface-to-surface maps $\Si\to M$.
Most of these maps are singular thus difficult to work  with. The main idea
of this work is to desingularize surface-to-surface maps by lifts
i.e. maps to an extended 
4-dimensional space-time $M\times M_v$.  The 2D space $M_v$ will be called the
vertical space.
It is known that the space of maps of  a 2D manifold to a 4D manifold
consists mainly of  non-singular maps called immersions.
The process of lifting is not unique. We shall  be able
to classify all possible lifts by a topological invariant.

Let us begin with the definition of an  immersion.
A map $X:\Si^s \to {\cal M}^m$  is an immersion if $rank(dX)=s$ i.e.  
the tangent map is of maximal possible rank ($m,s$ denotes dimensions of the
spaces, $s\leq m$).  
Roughly speaking it means that the image of $\Si^s$ in ${\cal M}^m$ is smooth.
It means also that the induced metric $g_{ab}\equiv \p_a{\vec X}\p_b{\vec 
X}$ is non-singular.
Any map $\Si^s\to {\cal M}^m$ can be approximated by an immersion  
for $2s\leq m$ \cite{whitneysi,dubrovin}. 
An immersion can have double points i.e. for some $p\neq q\in
\Si,\;X(p)=X(q)$. In the rest of this section we shall consider 
orientable ${\cal M}^{2s}$ of dimension $2s$, $s$ even. Hence, generically,
double points are isolated  and there are no triple points
(with an obvious definition). Moreover, the image of $\Si^s$ intersects
transversally i.e.  
$dX(T_p \Si^s)\oplus dX(T_q \Si^s)=T_{X(p)}{\cal M}^{2s}$ where $dX(T_p \Si^s)$ is the
image of the 
tangent space $\Si^s$ at $p$ under $dX$. In this case one defines the  
self-intersection number $I$  to be  equal to the 
number of double points summed with appropriate signs. 
The latter are chosen in the following manner: 
$(+)$ if $\{dX(T_p \Si^s),dX(T_q \Si^s)\}$ has the same orientation as
$T_{X(p)}{\cal M}^{2s}$ and $(-)$ in the opposite case.

There is a kind of homotopy  (regular homotopy) defined in the space
of immersions.
A regular homotopy is a homotopy which stays to be an immersion for each value
of the 
homotopy parameter.  Whitney \cite{whitneysi} showed that
the intersection number $I$ is invariant under  regular homotopies, thus in
this 
sense it is a topological invariant. Moreover  $I$ can be an
arbitrary integer number \footnote{It is defined mod 2 for non-orientable
$\Si^s$} and  
for immersions in the Euclidean spaces ${\cal M}^{2s}=R^{2s}$, $I$ 
classify all possible (up to regular homotopies) immersions \cite{smaleh}.
In general,
immersions $X:\Si^s\to {\cal M}^m$ are classified by monomorphism of tangent bundles
$T(\Si^s)\to T({\cal M}^m)$ \cite{smaleh,adachi}. The monomorphism is a bundle map
which 
restriction to each 
fiber of $T(\Si^s)$ is a vector space monomorphism i.e. for $(v_1,...,v_m)\in
T_p({\cal M}^m), \; dX: (v_1,...,v_m)\to (dX(v_1),...,dX(v_m))$ is a monomorphism.
It is clear that the key 
role in the classification is played by the space of  monomorphisms
$T_p(\Si^s)\to T_{X(p)}({\cal M}^m)$. If both manifolds have
Euclidean signature this  is the space of
$s$ linearly independent, orthogonal vectors in $R^{m}$ which is  
so-called Steifel manifold $V_{m,s}=O(m)/O(m-s)$. 
Inequivalent
immersions  $S^s\to R^m$ are related to elements of $\pi_s(V_{m,s})$.
For $M=S^2,\;N=R^4$ $\pi_2(V_{4,2})=Z$ and elements of $\pi_2(V_{4,2})$
are self-intersection numbers $I$. 
The analytic expression for  $I$  can be
easily found out in the literature \cite{whitneysi,lashof,mazur,ja-spin}. 
First one defines the normal bundle associated to a given immersion as bundle
of orthonormal vectors orthogonal to the image of $\Si^2$ in ${\cal M}^4$
under  $X$. 
Its structure group is $SO(2)$. Let the two normals be
$\{\vn_1,\;\vn_2\}$. Define a one-form $A=\vn_1 d\vn_2$ and its curvature $F=dA$. 
Then 
\beq
I[X]=\half c_2(n)=\frac{1}{4\pi}\int_\Si F
\label{chern}
\eeq
i.e. it is half of the Chern number (Euler number) $c_2(n)$  of the normal bundle.
In the  subsection (\ref{sec:riem}) we shall generalize this formula for
manifolds 
with boundary. 

\section{ A non-backtracking particle in 1D space-time. }
\label{sec:1d}

In this section we shall consider a 
simple 1D model of a massive non-backtracking particle. The example will
illustrate   basic features of the string case of interest.

In the first quantized language the dynamics of a particle moving in 1D
target space-time 
is given by the ordinary path integral with the action
$S[X]=\mu\int_0^1 dt \sqrt{\Xd^2}$, where
$X=X(t)$ is a map from the unit interval $[0,1]$ 
into the real ax $X\in R$. 
The set of paths contributing to  the path integral
contains  backtracking paths i.e. paths  for which
the velocity $\Xd$ 
changes sign. Points where $\Xd=0$  will be called singular 
and denoted by $P_i$ ($i$ enumerate such points). At $P_i$'s the map $X$
ceases to be an immersion.
This is a 1D model of a
fold discussed in  Sec.\ref{sec:r2r2}. 
We assume that points $P_i$ are isolated, what is the generic case.

We shall modify the action in such a way that  backtracking paths
will not contribute to the path integral. The idea is to
introduce another field $Y\in R_v$ (subscript $v$ means vertical), which 
will be interpreted as an additional (vertical) coordinate. The map
$(X(t),\;Y(t))$ is the lift of $X(t)$. We want to stress that although $X(t)$
maybe singular its lift, generically is an immersion.
We supplement the action  by a topological term $R$:
\beq
S[X,Y]=\int_0^1 dt \mu\sqrt{\Xd^2} +i\pi R
\label{lengthn}
\eeq
where $R=\frac{i}{2\pi} \int_0^1(\Xd \Ydd-\Yd \Xdd)
/(\Xd^2+\Yd^2)$ (for flat space-times) is the
rotation number  of the vector tangent to the path $(X(t),\;Y(t))$.
In the parameterization  $\Xd=v \cos\a,\; \Yd=v \sin\a\;$, 
$R=\int_0^1 dt {\dot \a}(t)$. This formula gives a unique  answer if
boundary values of $\a$ are fixed. Thus we set
$Y(0)=Y(1)=\Yd(0)=\Yd(1)=0$ at the ends of paths or 
$Y=Y_0,\;\Yd(0)=\Yd(1)$ for closed paths.
All maps $(X,Y)$ we
sum over under the function integral are immersions. This is a dense set of
maps in the space of smooth maps.

The model (\ref{lengthn}) has an enormous group of 
local symmetries. These are local reparameterizations  and 
$v$-regular homotopies (VRH) $\d_y Y(t)=\ep(t)$,
where $\ep$ is  such that ($X, \;Y+\ep$) is an immersion.
Below we describe topological sectors of the model (\ref{pathn}) i.e. lifts 
which differ only by a VRH.
Let the path $X(t)$ has folds (see Fig.1a),
the latter are characterized by positions of singular points $P_i$ 
 : $\Xd(P_i)=0$ for all $i$. 
Because we work with immersions, $\Yd(P_i)\neq0$ necessarily. Thus either
$\Yd(P_i)>0$ or $\Yd(P_i)<0$, because  VRH can not change the 
sign of $\Yd(P_i)$. We can assign  the following set of maps
to any immersion $f_i:\;P_i\to 
{\rm sign}(\Yd(P_i))$  (all $i$).
We shall show that {\it any two lifts 
are $v$-regularly homotopic (VRH)  if and only if
they correspond to the same  set \{$f_i$\}}.
If two different lifts $(X(t),\,Y_1(t))$,
$(X(t),\,Y_2(t))$ are VRH then we can make
$Y_1=Y_2$ what defines \{$f_i$\} uniquely. One the other hand let us assume that
two immersions are characterized by the same \{$f_i$\}.  Using
a VRH one can set $Y_1(P_i)=Y_2(P_i)$. 
Thus, both immersions are equal
in the infinitesimally small  neighborhood of $P_i$'s (up to second power
of an infinitesimal quantity) because \{$f_i$\} gives equality of tangents to 
both immersions at $P_i$'s. 
Away from the singular points $X(t)$ is an immersion. In this case the
shift parameter $\ep (t)$ defined by a VRH can be arbitrary 
so one can make $Y_1(t)=Y_2(t)$ there. Thus one can do it everywhere.
We conclude that \{$f_i$\} uniquely defines topological sector of maps. 
\begin{figure}[t]
\label{pathnf}
\vspace{0.5cm}
\postscript{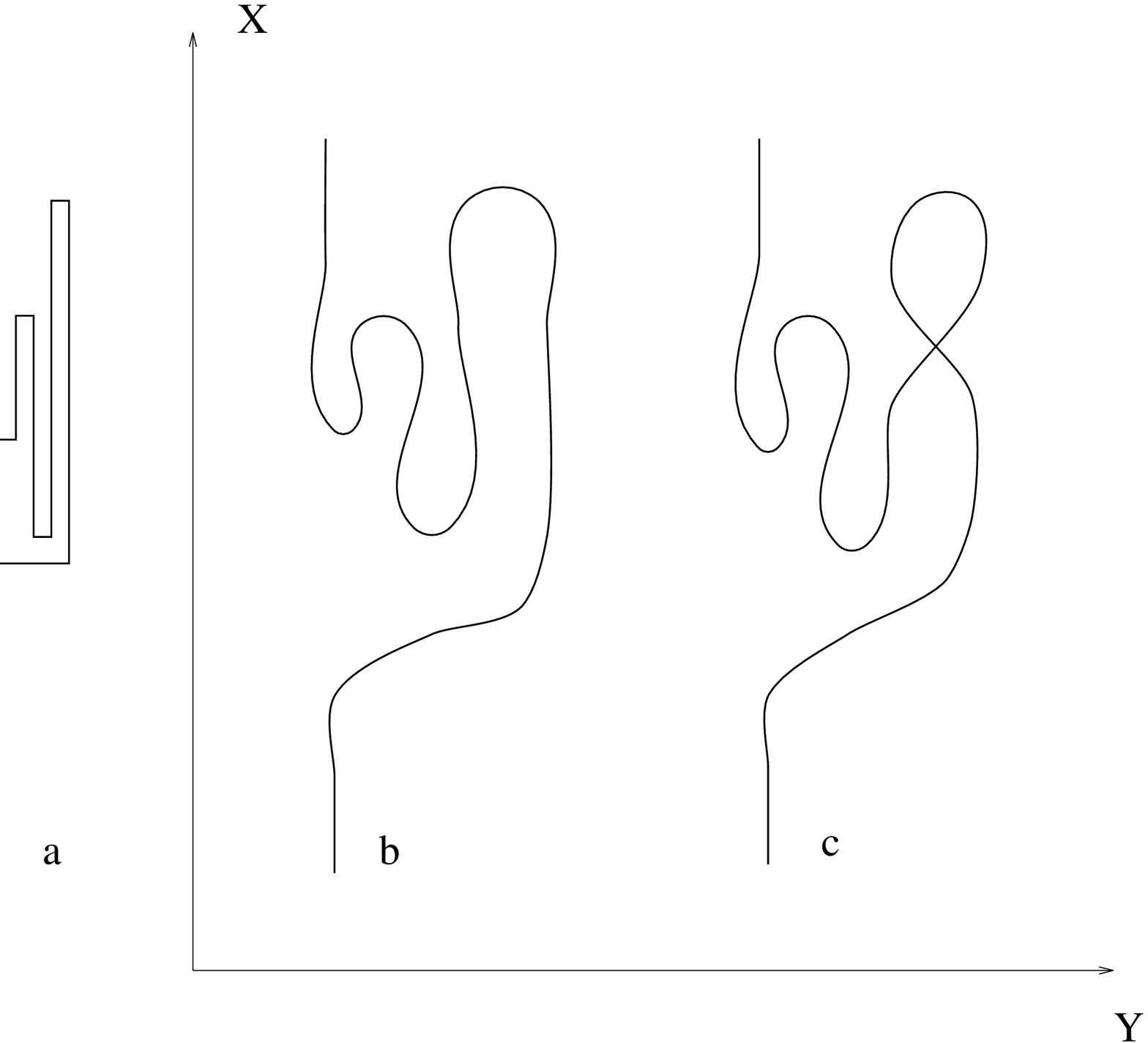}{0.57} 
\vspace{0.5cm}
\caption{(a) A one dimensional path $X(t)$ with  folds. (b,c) Two lifts of
the path (a) which 
rotation numbers  differ by one.}
\end{figure}

We give a heuristic  definition of the functional 
measure. Formally the measure is defined through:
\beq
\frac{\cD X}{Vol(Diff)}\frac{ \cD Y}{Vol({\rm VRH})}
\eeq
where $Vol(Diff)$ is the volume of path diffeomorphisms and $Vol({\rm VRH})$ 
is the volume of VRH's. First we note that the space of VRH's
depends on $X(t)$, because 
VRH's are defined by $\d_y Y(t)=\ep(t)$,
where $\ep$ is  such that ($X, \;Y+\ep$) is an immersion. 
On the other hand the VRH's
do not depend on a topological sector of the 
model i.e. for any choice of \{$f_i$\} the set of allowed 
VRH's is the same. Thus we can write
the following  
expression for the measure:
\beq
\frac{\cD X}{Vol(Diff)}\frac{ \cD Y}{Vol({\rm VRH})}=
\frac{\cD X}{Vol(Diff)}F[X]\sum_{\{f_i=\pm 1\}}
\eeq
where $F[X]$ is a, possible trivial, functional of $X(t)$. 
The sum runs over different 
lifts for given singular point $P_i$. 
The cancellation 
we are going to describe is due to this sum and is independent on $X$.

After  gauging away the VRH the path integral for
the  model is
\beq
\int \cD X e^{-S[X]}F[X]\sum_{\{f_i=\pm 1\}}e^{i \pi R[\{f_i\}]}
\label{pathn}
\eeq
It is easy to see that different lifts differ by rotation numbers.
An example of a map with folds and two lifts is presented in Fig.1.
Lifts which differ at one singular point $P_i$ have
 rotation numbers different by 1 (see Fig.1). Hence their
contributions  cancel out under the functional integral (\ref{pathn}). 
The sum $\sum_{\{f_i=\pm 1\}}$ is performed independently for each $i$ thus
cancelation holds for any set of $P_i$'s.
Thus all folds will be suppressed from
the functional integration (\ref{pathn}).
We are left with paths without folds i.e. such that the velocities during
the propagation do not change sign. The best way to define the theory without
folds is to localize the functional measure on a
specific set of paths \cite{moore,horava}.  This fixes the theory (e.g. $F[X]$) 
unambiguously. 

Below we calculate the propagator of the 1D non-backtracking
particle i.e. the  transition 
amplitude from $X_0=X(0)$ to $X_1=X(1)$ on the straight line $X\in R$.
Boundary conditions for $Y$ are fixed  as it has been discussed above.
Due to trivial topology of the $X$ space and the reparameterization symmetry
only one path contributes. It corresponds  to
the motion with the constant speed. Thus the transition amplitude is
$e^{-\mu l}$, where $\mu$ is the mass of the particle and $l=|X_1-X_0|$.
If $X\in S^1$ then there are also paths winding around the circle.
Paths winding $n$ times have the rotation number equal $n$.  
One can see it 
mapping the
cylinder $S^1\times R$ on the  plane $R^2$ and then calculating the
rotation number as in (\ref{lengthn}).
This  leads to the following expression for the propagator:
\beq
\frac{1}{1+e^{-\mu L}}\lbr \theta (l)e^{-\mu l}-\theta (-l) e^{-\mu (l+L)}
\rbr
\label{circle}
\eeq 
where $L$ denotes the circumference of the circle and
$l$ is within $(-\half L,\half L)$ . The factor 
$1/(1+e^{-\mu L})$ appears due to the infinite summation over
winding sectors with the alternating term $(-1)^I$.
The relative sign $-1$ of both contributions comes also from $(-1)^I$ :
paths contributing to the second term of (\ref{circle}) have 
rotation numbers different by one (they are accompanied by the
additional  $e^{-\mu L}$).

The model (\ref{pathn}) is in fact a certain
compactification of the 2D fermionic particle \cite{polbook}.
If we think of the vertical direction $Y$ as being compactified then its
topology 
should be that of $S^1$ instead of $R^1$.
Proposed above scheme  works also in this case but the arguments are
a bit more involved. There are infinitely many 
topological sectors for
each singular point due to winding modes
around $S^1$. Instead of two lifts we get two sets of
lifts.
Anyway the cancellation holds in a similar manner.

\section{String without folds.}
\label{sec:string}

The string theory functional integral for 2D targets is a sum over 
surface-to-surface maps $\Sigma\to M$, where $\Sigma$ denotes the string
world-sheet and $M$ the 2D target space-time. Surface-to-surface maps were
describe in Sec.\ref{sec:r2r2}. Generic map  have singularities: folds and 
cusps. 
As works on QCD$_2$  and YM$_2$ shows  singular maps do not
contribute to physics of these theories \cite{bars,gross}. Thus we need a
mechanism to  
remove their contribution.
Here we will follow the lines of reasoning of Sec.\ref{sec:1d}. In direct
analogy with 
this section we shall introduce a 
string model and show that
for flat target space-times the  model suppress generic i.e. folded contributions.
What about non-generic 
surface-to-surface maps? For flat space-times there are no non-singular maps
and because the space of maps with folds and cusps is
dense in the space of all smooth maps \cite{r2r2} one can safely claim that
these maps also do not contribute. Things get more complicated if target
space can be arbitrary Riemann surface. Then there are maps without 
singularities and these can not be simply discarded. We shall discuss these
problems in subsections \ref{sec:riem}.

\subsection{String model}

In this subsection we introduce a string model which suppresses folds.
In strict analogy with Sec.\ref{sec:1d} we introduce
two additional, (vertical) world-sheet fields: $(Y^1,\;Y^2)\in
M_v$ ($M_v$ is a  ``vertical'' surface). The functional integral is over
$(X^1,\;X^2,\;Y^1,\;Y^2)\in M\times M_v$ configurations. The latter can be
viewed as lifts of $(X^1,\;X^2)$ configurations and we know from
Sec.\ref{sec:imm} that generically they are immersions.
The proposed string action is a direct generalization of the particle case
(Sec.\ref{sec:1d}). 
\beq
S[X]=\mu\int_M d^2\si\sqrt{g}+i\theta I[X,Y],
\label{arean}
\eeq
$I$ is the self-intersection number of the  surface immersed 
in the 4D space \cite{whitneysi,lashof,smaleh,polrig,mazur,ja-spin}.  
The vertical coordinates  enter the action only through $I$.
The action (\ref{arean}) is invariant
under arbitrary VRH of the vertical fields \cite{whitneysi}:
$\d Y^\mu(\xi)=\ep^\mu(\xi),\;  (\mu=1,\; 2)$.

Cancellation of folds for $M=M_v=R^2$ was presented in Ref.7.
Its basic ingredient is the classification of lifts and the alternating sum
provided by the self-intersection number $I$. Below we shall
generalize these results for arbitrary $M$ and
$M_v$.

\subsection{Classification of lifts. }

In the following we  are going to classify  topological sectors of the
model.  
We say that two immersions are in the same topological sector if they can
be connected by a VRH.
In the following $X$ will denote a map $X:\Si\to M$  with folds and $(X,Y)$
its lift 
into the extended 4D space-time:
$(X,Y):\Si\to M\times M_v$, where $M_v$ is an arbitrary 
orientable  surface without boundary. In Sec.\ref{sec:r2r2} we  
defined the line bundle $K$ bundle by $dX(K)=0$, where  $dX$
is taken at  
 points belonging to $S_1$ and the tangent map acts on the fiber over 
that point. In our case $K$ is trivial bundle ($w(K)=0$).
Lifts of the fold must have non degenerate 2D tangent space, hence 
must respect $dY(K)\neq 0$. In this way the couple $(Y,dY)$ defines a map
(monomorphism) 
from $K$ to the tangent bundle of $M_v$ ($TM_v$). 
The set of connected components of such maps,  define different topological
sectors of lifts. The condition  $dY(K)\neq 0$ imply that this
is the same as the set of connected 
components of maps from $K$ to one dimensional sphere bundle over $M_v$ 
($SM_v$).
For one fold it is 
given by $\pi_1(SM_v)$, because the fold has topology of $S^1$. We conclude that
$\pi_1(SM_v)$ classify lifts of one fold \cite{smale}. Generalization for the
case of a  
map with many folds is obvious, because lifts of folds are independent on 
each other. 

As an example \cite{ja-fold} we take first $M=R^2=M_v$.
Then lifts of the i-th fold are classified by $\pi_1(R^2\times S^1)=Z$.
The integer $f_i\in \pi_1(R^2\times S^1)$ is invariant under the 
VRH and is directly
related to the self-intersection number $I$ of the lifted configuration. 
We can see it if we notice that both numbers are additive under
gluing \cite{ja-fold}.
Then the  self-intersection number is
$I[f]=\sum_{folds}\pm f_i$. \footnote{ Strictly speaking 
the last formula holds if $I=0\Leftrightarrow
f=0$ holds what is true if  $w(K)= 0$ (see Sec.\ref{sec:r2r2}). For
$w(K)\neq 0$, 
$f=0$ may not correspond to $I=0$, because then a surface-to-surface map
can not be lifted into $R^3$ \cite{blank}.   
In this case the formula for $I$ may differ by an additive constant (an
integer) which 
is unessential for cancellation of folds (see (\ref{sum})). This more
complicated case was 
discarded in Ref.7 for the  sake of simplicity.}

Now we go to the string theory. We want to show that the originally folded
 configurations $(X^1, X^2)$ will cancel out from the partition
function. Here one should discussion the 
construction of the 
functional integral measure of the theory. The arguments are straightforward
generalization of that given in Sec.\ref{sec:1d} thus we omit them here.
We get the following expression
for the functional integral:
\beq
\int \cD X^1 \cD X^2\;e^{-S[X]}F[X]\sum_{\{f_i\}}e^{i \theta I[f]}
\label{surfacen}
\eeq
The sum over $f_i$'s can be performed independently for each $i$ because
$I[f]=\sum_{folds}\pm f_i$. For one fold we get 
\begin{equation}
\sum_{f\in Z}e^{\pm i\theta f}=2\pi\delta (\theta)
\label{sum}
\eeq
Thus all folded configurations vanish from the path integral for 
non-zero $\theta$.
Maps contributing to the  vacuum-to-vacuum amplitude of the closed string 
necessarily have folds for the target space $R^2$. According to the
above discussion the amplitude vanishes. This also holds for any correlation 
function of any finite set of local operators.
Thus the final conclusion of this part of the paper is that 
the model (\ref{arean}) is trivial for the $R^2$ space-time.

The model discussed in this subsection contained one addition parameter 
compared to YM$_2$ with semisimple Lie group: it is the $\th$ angle \footnote{
I would like to thank I.Kogan for discussion on this point.}. 
In the following subsection we shall claim that this $\th$ is 
inherited from YM$_4$.

\subsection{Compactification}

One can view (\ref{arean}) as a certain compactification of 
a 4D string. Thus $M_v$ is  a compact manifold without boundary.
If a characteristic size of this space is small we expect that 
quantum fluctuations in the compactified directions are strongly suppressed.
The standard compactification of YM$_4$ on a 2D torus leads to 
additional 2D degrees of freedom: the adjoint matter. Its because the 4D
gauge fields $A^a_\mu(x^1,\dots x^4)$ ($\mu=1,\dots 4,\;a$ is the adjoint
representation index) decompose into 
$(A^a_\a(x^1,x^2),\; A^a_3(x^1,x^2)$, $A^a_4(x^1,x^2))$ ($a=1,2$).
From the point of view of the uncompactified 2D space-time $A^a_3,\;A^a_4$ are
the matter fields. 
The appearance of the continuous $\theta$ parameter in (\ref{sum}) is
natural in this case. It may correspond to the analogous angle in the QCD 
Lagrangian, because after compactification we have:
\beq
\int d^4x tr(F{\tilde F})\to 4\int d^2x \ep^{\a\b}tr(F_{\a\b}A_3A_4)
\eeq
In order to get rid of these matter fields (an thus certain ambiguities in the
construction) we shall compactify on $S^2$.

Topology of the vertical space significantly changes  the classification of lifts. 
Instead of Z inequivalent regular homotopy sectors we obtain
only two. The general arguments presented in the previous subsection say that 
lifts of one fold are classified by $\pi_1(S(S^2))=Z_2$ \cite{smale}. 
The trivial element of this group corresponds to $I=0$,  the  non-trivial
elelment to $I=1$.
Now the cancellation of folds holds only for $\theta=\pi$, 
as for the particle case.
In this way we fixed the value of the only free parameter in the model. It is
in accordance with YM$_2$ which for semisimple Lie groups 
do not have any angle-like parameter.

\vs

\subsection{$M$ as a Riemann surface}
\label{sec:riem}

In this subsection we shall consider shortly the case when $M$
has non-trivial topology e.g. it is a Riemann surface. First we briefly
recapitulate Gross/Washington interpretation of the 1/N expansion of the
$SU(N)$ partition function. The model almost decouple into two (chiral)
sectors: orientation preserving and orientation reversing maps. Both sectors
are connected only by orientation reversing infinitesimally thin tubes 
(${\tilde t}$). One chiral sector looks like branched cover with simple branch 
points $(i)$. Additionally there are orientation preserving infinitesimally
thin tubes ($t$) connecting different sheets of the branched cover and 
collapsed handles ($h$). We stress again that $({\tilde t})$, $(i)$, ($t$),
($h$) maps have singularities of the non-generic type.

Let us note that  cancellation of folds holds
for arbitrary $\Si$, $M$, $M_v$. It is
due to additivity of $I$ under glueing.
But this
is not the whole story: contrary to $M=R^2$ case the
space of maps without folds is not empty. Although these maps are not generic
we do not have any argument to dismiss them.
Maps which do not have singularities obviously have unique lifts.
Brenched covers have have $2^i$ lifts ($i$ is the number of branching points).
The question is what $I$ have their lifts ?
Locally a map with a (simple)
branch point at $z=\si+it=0$ looks like $z\to X=z^2$ for $X,z\in C$ and
${\rm   rank}(dX|_{z=0})=0$ i.e. $dX$ has two null eigenvectors. 
It follows that any lift must have two lineary independent tangent
vectors at $z=0$ lying in the vertical directions, thus forming an area
element of  $M_v^2$. This implies that  a branch point can be lifted in two ways
differing the orientation of the area form:
$$
\{z\to z^2\}\stackrel{lift}{\to}\{z\to (z^2,z) \; {\rm or}\;(z^2,{\bar
z})\}
$$ 
The important observation is that branched covers  
represent  elements of the second homology group $H_2(M)$ and their lifts
elements of $H_2(M\times M_v)$.
Therefore we can calculate its algebraic self-intersection
number defined as below \cite{bredon}:
$$ 
a\cdot b=D^{-1}(D(a)\wedge D(b)),
$$
where $D$ denote the Poincare duality
(isomorphism), $ D:H_2(M\times M_v)\to H^2(M\times M_v)$ and $a,b\in
H_2(M\times M_v)$.  
Vanishing of $a\cdot a$ implies that  there  is an embedding 
of $a$ in $M\times M_v$ \cite{kervaire} i.e. $I=0$. 
If $H_2(M_v)=0$ then $H^4(M\times M_v)=0$ so $a\cdot a=0$ always.
In oreder to simplify consideration for the case $H_2(M_v)\neq 0$  we  take
lifts that do not 
wind around $M_v$. Then $D(a)\in H^2(M)$, so $a\cdot a=0$. 
In all cases we get $I=0$ so the self-intesection term does not influence
the functional integral. For example it does not change results of
Ref.5 if we 
treat this work as a realization of (\ref{arean}).

In the end we add few comments about other  maps which
contribute to the YM$_2$ partition function. Both infinitesimal tubes $t$ and 
${\tilde t}$ have
singularities along a circle $S^1$. 
Tubes $t$  can be viewed as a collision of two branch 
points \cite{moore}. If so they should be lifted with $I=0$. 
We can not reach this conclusion from the analysis of lifts alone. 
In fact there are Z lifts of $t$ for $M_v=R^2$ because the circel of 
singularities can be mapped to an arbitrary
 smooth loop on $M_v$. 
The situation for  tubes ${\tilde t}$ is  analogous.

We see that the simple approach proposed in this model can not clarify all
features of Gross-Taylor interpretation \cite{gross}. No wonder it is the
case.  
This goal can be partially achieved by application of more subtle machinery:
topological field theory methods. The 
precise description of one chiral sector of the 
model can be achieved if the
topological theory is localized on holomorphic maps (branched
covers) \cite{moore}. Tubes  
$t$ were obtained by a compactification of this space of maps. It is known
that the process of compactification is not unique. The other compactifications
presumably corresponds to generalized YM theories \cite{witten,ya}.
Despite these successes the contribution of  ${\tilde t}$ tubes
have not been 
completed. Moreover the extension to 4D target space-times is not clear.
In this palace we would like to stress that the approach proposed in this paper
shows {\it why we should localize on a space of holomorphic or harmonic maps
(minimal area maps)}.

\subsection{4D theory  and string states }

It is clear that 
the model (\ref{arean}) has straightforward extensions to 3 and 4
dimensional space-times.  
It is enough to make the additional dimensions dynamical i.e.
add them to the Nambu-Goto action. Higher dimensional string models
may require more
terms e.g. the extrinsic curvature term \cite{polrig,rig}.  

In this subsection we shall dwell upon a problem of description of string states.
We define string states to be framed loops in $R^3$ \cite{mazur}.
Let $C$ be an oriented, closed curve   immersed  in $R^3$.
A framing of $C$ is
a field of unit  vectors $\vn$ orthogonal to $C$. Then
one defines: the (Gauss) linking number $L$, the twist $T$ and the writhe $W$ 
of the framed curve $C$ as follows \cite{white,grunberg}:
\beqa
L[C,C']&=&\frac{1}{4\pi}\int_C dX^i\int_{C'}dY^j\ep^{ijk}
\frac{(X^k-Y^k)}{|X-Y|^3}\\
T[C,C'] &=&\frac{1}{2\pi}\int_C d\si\ep^{ijk}\frac{\Xd^i}{|\Xd|}n^j\p_\si
n^k\\
W[C]&=&\frac{1}{4\pi}\int_C dX^i\int_{C}dY^j\ep^{ijk}
\frac{(X^k-Y^k)}{|X-Y|^3}
\label{link}
\eeqa
where $C'$ is  a parallel shift of $C$ by $\ep\vn$, where $\ep>0$ is an
infinitesimal parameter. Using  ${\vec Y}={\vec X}+\ep\vn$ for (\ref{link})
one can easily 
show that : $L[C,C']=W[C]+T[C,C']$ (White formula). It follows
that for flat 
curves the writhe is zero thus the linking number equals the twist.
The writhe changes by 2 if a curve self-intersects. In order to show it we take
two curves: $C$ which nearly self-intersects and  $C'$ which is a small
deformation of  $C$ such that $C'-C\equiv C_\ve=\p S$ and the surface $S$ 
intersects $C$ (see Fig.2). 
\begin{figure}[h]
\label{writhf}
\vspace{0.5cm}
\postscript{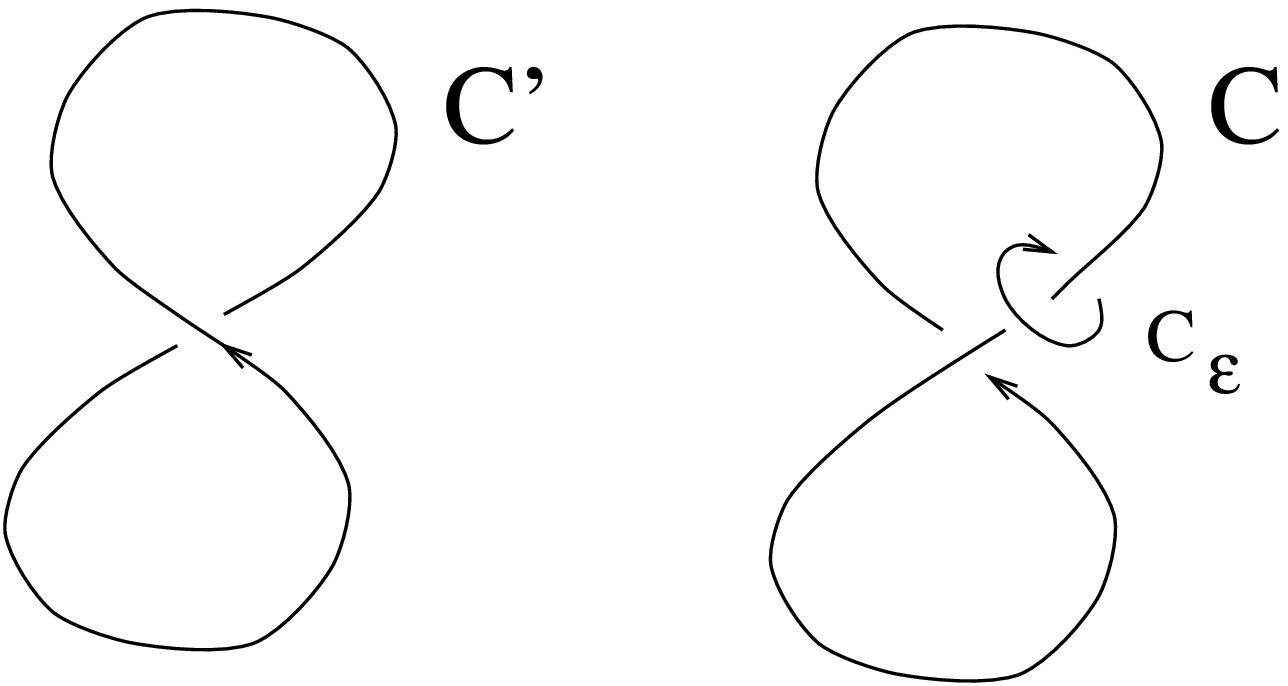}{0.50} 
\vspace{0.5cm}
\caption{$C'-C\equiv C_\ve=\p S$. The
small circle around $C$ is $\p S$.}
\end{figure}
We take $S$ to be flat. Then directly from 
the definition of $W$ we get: $W[C']-W[C]=2 L[C_\ve,C]+W[C_\ve]$. When
$C_\ve$ shrinks to zero
$L[C_\ve,C]=1$ (with orientation as in Fig.2) and $W[C_\ve]=0$.
Thus $W[C']-W[C]=2$.

Let the target space ${\cal M}^4$ be a 4D contractable manifold with boundary $\p {\cal M}^4
$, $\Si$ a 
Riemann surface with boundary $\p \Si$ and $X:\Si\to {\cal M}^4$ an immersion such that
$X(\p\Si)\subset\p {\cal M}^4$. We take ${\cal M}^4=R^3\times 
D^1$ ($D^1=(t_i,t_f)$ is a time  period), so $\p {\cal M}^4=R^3\cup R^3$.
Under these conditions we
can write down a formula for the self-intersection number of surfaces 
with boundary.  It is given by:

\beq
I_{\Si,\p\Si}[X]=\half (c_2(n)-W[X(\p\Si)])
\label{bound}
\eeq
In  (\ref{bound}) the writhe should be calculated with the proper orientation of
$\p {\cal M}^4$, because $W$ changes sign with change of orientation.   
$I_{\Si,\p\Si}[X]$ is invariant under those regular homotopies of $X$ which do
not lead to self-intersections of the boundary.  The proof of
the formula is very simple. Take a deformations $\d X$ of $X$. If $\d
X$ has support inside $\Si$ then $\d c_2=\frac{1}{2\pi}\int_\Si d\d A= 0$ 
and $\d W=0$ trivially. 
Now let $\d X$ has support which includes $\p \Si$.
Because $X(\p\Si)\subset\p {\cal M}^4$ both normals
are tangent to $\p {\cal M}^4$. Let $\si$ parameterize ${\vec X}(\p\Si) $ and 
$\{\p_\si {\vec X},\vn_1,\vn_2\}$ be positively oriented then
$\vn_2\propto(\p_\si {\vec X}\times \vn_1)$.
Hence we get the following sequence of 
equalities:  
$\d (c_2(n)-W[X])=\frac{1}{2\pi}\int_\Si d\d A-\d W=\frac{1}{2\pi}
\int_{\p\Si}\d(\vn_1 d\vn_2)-\d W=
-(\d T+\d W)=-\d L=0$. The last equality holds because linking number is
invariant under deformations which do not produce self-intersections.
If the deformation produces a self-intersection then $W$ changes by 2 (with
an appropriate sign convention) changing $I$ by 1.

If a closed string propagats in the 4D Minkowski space-time the
topology of the world-sheet is $S^1\times D^1$ ($D^1=(t_i,t_f)$ is a 
time  period). Immersions must map
the world-sheet time to the target time-like direction. 
It is easy to see that
they are classified by immersions of $S^1\to R^3$ which are trivial because
$\pi_1(V_{3,1})=\pi_1(S^2)=0$. 
Hence the normal bundle is trivial and we can take $F=dA$, with globally
defined $A$. We define string states as non-self-intersecting framed loops.
Any such a state carry an
integer quantum number: the linking number $L\in Z$ defined as above.
For two such states $|L_1\ran, |L_2\ran$ the Minkowski surface connecting them
must self-intersects $(L_2-L_1)$ times. One can see it noticing that because
of triviality of the normal bundle $I_{\Si,\p\Si}[X]=1/2(\int_{\p
\Si}A-W)=L_2-L_1$. 
Thus if a string action contains a $\theta$-term 
($\theta I_{\Si,\p\Si}$) the transition amplitude $\lan L_2|L_1\ran$ would be
proportional to $exp\{i\theta (L_2-L_1)\}$. 
It was climed that   
the self-intersection number $I$ plays similar role in 4D string theory as 
$F{\tilde F}$ in gauge theory \cite{bal,mazur,ja-spin}. 
If so one sees the full correspondance between framed string states and
different topological 
sectors of QCD.

\section{Application to 3D YM theory}
\label{sec:3d}

We expect that 3D gauge theories also have a string representation. 
Such  systems are interesting by
itself but also provide a model for the critical behaviuor of the 3D Ising
model about which we comment in the end of this section.
In analogy
with 2D Yang-Milles theories one might expect that also 3D theories  
suppressed singular configurations.

Let us briefly describe the space of maps of a surface $\Si$ to 3D Euclidean
space $R^3$. A generic map is not an immersions but poses singularities: the
so-called cross-caps \cite{cross-caps} (see Fig.3), where 
the tangent map drops rank by one. Generically, crosscaps are points $P_i$.
\begin{figure}[h]
\vspace{0.5cm}
\postscript{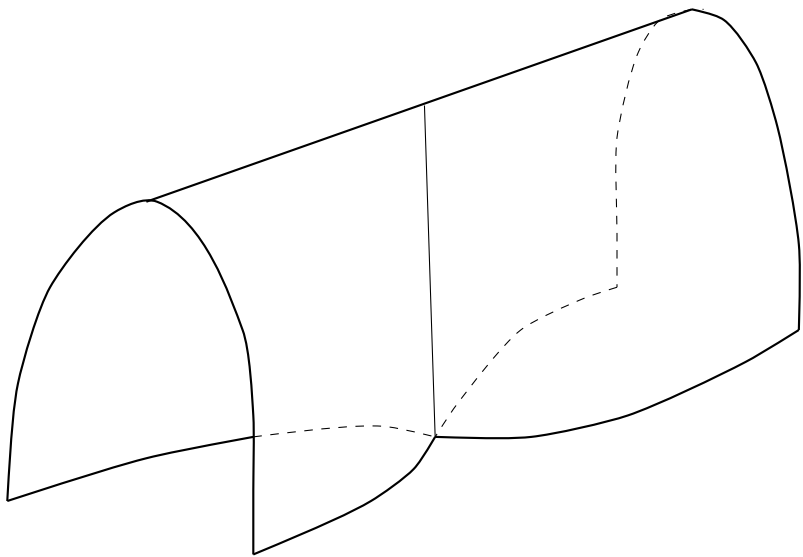}{0.67}              
\vspace{0.5cm}
\caption{A cross-cap.}
\label{cross-cap}
\end{figure} 
Any cross-cap ends a line of
self-intersections. There is a coordinate system in which the cross-cap has
the following form:
\beq
(\si,t)\to X=(\frac{t^2}{4}, \si,\frac{\si t}{2})
\label{ccap}
\eeq
Here the map $X$ is singular at one point $P_1=X(\si=0,t=0)$, where the vector
$\p_\si X$ is zero. Fig.3 shows the neighborhood of $P_1$.
In analogy with 2D Yang-Milles theories one might expect that also 3D theories
suppressed singular configurations. One can
do it appling a mechanism which is
basically  the same as that of  Sec.\ref{sec:string}. 

We extend the space $R^3$ to $R^4$ and consider the space of immersions
$\Si\to R^3$. 
As a string  action we take $S^{(4)}=S^{(3)}+i\pi I$, where $S^{(3)}$ is a 3D
action. $S^{(4)}$ is invariant under VRH's but now the vertical space is 
one dimensional just as in the 1D particle case, hence the  classification  of
all lifts to  $R^4$ is the same i.e. they are characterized by ${\rm
sign}(dX^4(P_i))=Z_2$. 
It means that to each 3D singular map there corresponds two inequivalent 4D
lifts. Their
self-intersection numbers differ by one so their contributions will cancel
out from the path integral. We show it for a map $X$ of the sphere
$X:\Si=S^2\to R^3$ having one line of self-intersections ending with two
singular points $P_1,\; P_2$. 
If ${\rm sign}(dX^4(P_1))={\rm sign}(dX^4(P_2))$ the lift has $I=0$. We can see it
constructing a homotopy of $X$ which glue $X(P_1)$ with
$X(P_2)$. In this case it is a regular
homotopy because the tangent map is contineaus and nowhere vanishing. Contrary, for
${\rm sign}(dX^4(P_1))\neq {\rm sign}(dX^4(P_2))$ one can not construct such a
regular homotopy so $I\neq 0$. 
As a convenient realization of the situation we
take the following map:
\beq
X=(\si,t-\frac{a\,t}{(1+\si^2)(1+t^2)},\frac{\si t}{(1+\si^2)(1+t^2)})
\label{map}
\eeq
It has two singulat points for $a>1$ which join each other when $a\to 1$ and
finally vanishes for $a<1$.
If $X^4=t/[(1+t^2)(1+\si^2)]$ then we have the former situation and one can
explicitely calculate the self-intersecton number, $I=0$ . For 
$X^4=1/[(1+t^2)(1+\si^2)]$ the self-intersection number is $I=-1$.
This argument shows that changing ${\rm sign}(dX^4(P_i))$ changes $I$ by one.
Thus including an alternating factor $e^{i\pi I}$ and counting maps as in
Sec.\ref{sec:string} we suppress all  maps with cross-caps. Thus we are left with
immersions into 3D space. 
It is worth to note that the mechanism
works for any 3D action e.g. $S^{(3)}$ may contain the extrinsic curvature.

Finally we comment on the 3D Ising model.
In the dual picture the 3D Ising model is the 3D $Z_2$ gauge theory
 \cite{savit}. 
It is  known \cite{david} that a gas of self-avoiding
surfaces in a 
spatial 3D lattice is in the same universality class as the Ising
model. Surfaces we are talking about bound bulks of 3D space in which
spins are oriented in the same direction. One expects that the energy of
a spin  configuration is proportional the area of these surfacee i.e. we
expect the action to be proportional to the 3D Nambu-Goto term.
The problem is how to describe self-avoiding surfaces in a string
theory language ? Self-avoiding surfaces are embeddings thus
for the 3D Ising model the space of immersions is too large. We
need to localize the functional integral on the space of embeddings. 
Unfortunately it is not known how to do it. A possible solution was proposed
few years ago \cite{distler,ja-ising} where the cancellation of self-intersecting
maps could hold due to summation over topologically different world-sheets.
Unfortunately not much has been done in this direction.
\vskip1cm 

{\bf Acknowledgment}. I would like to thank T. Mostowski for 
discussions concrerning geometrical aspects of the paper, 
K.Gaw\pole dzki, R.Dijkgraaf, I.Kogan for valuable comments
 and A.Niemi for kind hospitality in Uppsala University where a part of this 
paper has been written.

\end{document}